\documentclass[%
 reprint,
 amsmath,amssymb,
aps,
pra,
]{revtex4-2}
\usepackage{color,soul}
\usepackage{graphicx}
\usepackage{dcolumn}
\usepackage{bm}
\usepackage{natbib}
\usepackage{mathtools, nccmath}
\begin{document}

\newcommand{\beginsupplement}{%
        \setcounter{table}{0}
        \renewcommand{\thetable}{S\arabic{table}}%
        \setcounter{figure}{0}
        \renewcommand{\thefigure}{S\arabic{figure}}%
     }

\preprint{}

\title{Electropolishing-Induced Topographic Defects in Niobium: Insights and Implications for Superconducting Radio Frequency Applications}
\thanks{hryhoren@jlab.org}%

\author{Oleksandr Hryhorenko\,$^{1*}$, Anne-Marie Valente-Feliciano\,$^{1}$, and Eric M. Lechner\,$^{1}$}
 \affiliation{$^{1}$Thomas Jefferson National Accelerator Facility, Newport News, Virginia 23606}

\date{\today}

\begin{abstract}

Electropolishing is the premier surface preparation method for high-Q, high-gradient superconducting RF cavities made of Nb. This leaves behind an apparently smooth surface, yet the achievable peak magnetic fields fall well below the superheating field of Nb, in most cases. In this work, the ultimate surface finish of electropolishing was investigated by studying its effect on highly polished Nb samples. Electropolishing introduces high slope angle sloped-steps at grain boundaries. The magnetic field enhancement and superheating field suppression factors associated with such a geometry are calculated in the London theory. Despite the by-eye smoothness of electropolished Nb, such defects compromise the stability of the low-loss Meissner state, likely limiting the achievable peak accelerating fields in superconducting RF cavities. Finally, the impact of surface roughness on impurity diffusion is investigated which can link surface roughness to the effectiveness of heat treatments like low-temperature baking or nitrogen infusion in the vortex nucleation or hydride hypotheses. Surface roughness tends to decrease the effective dose of impurities as a result of the expansion of impurities into regions with greater internal angle. The effective dose of impurities can be protected by minimizing slope angles and step heights, ensuring uniformity.

\end{abstract}

\maketitle


\section{\label{sec:Intro} Introduction}

Superconducting radio frequency (SRF) technology is the primary means for high-duty cycle and CW operations, delivering high-energy ($\sim$ TeV), and high-current ($\sim$ mA) particle beams through cavities typically made of bulk Nb or Cu coated with a Nb film \cite{reece2016continuous, SingerReview2016, Bousson2014, Boussard1999, Gerigk2018, porcellato:epac04-tupkf024, Morozov2022,biarrotte2013, padamsee2014superconducting}. The performance of SRF cavities is fundamentally dependent on the quality of their inner surface hosting intense radio frequency (RF) waves.  Nowadays, electropolishing (EP) and buffered chemical polishing (BCP) are polishing techniques that provide reliable and reproducible field-dependent surface resistance. There are different advantages for each of the aforementioned polishing techniques, but the main advantage of EP over BCP is a smoother surface, which is an essential requirement for high-Q and high-gradient applications. The X-ray Free Electron Laser (XFEL) machine is the largest deployment of electropolishing dedicated to SRF applications to date, which has resulted in processing more than 800 superconducting cavities \cite{SingerReview2016}. The statistics collected show that a 2-step EP with a 150 µm total material removal can achieve an average gradient over 30 MV/m at quality factor $\geq 10^{10}$ \cite{SingerReview2016}. The cause for some low performing cavities could be traced back to the defects in the cavity inner surface. Such defects can be improved by a combination of mechanical polishing (local grinding or centrifugal barrel polishing) with EP or buffered chemical polishing \cite{Cooper2013MirrorSmooth, yamamoto:ipac12-weppc013, massaro:srf15-mopb094}. In the absence of such obvious defects, what limits the achievable peak magnetic fields within Nb SRF cavities? 

Performance in superconducting radio frequency cavities is inherently complex and multifaceted. Along one of these facets, preservation of the low-loss Meissner state is critically important. The field at which the Meissner state breaks down is given by the superheating field. At the superheating field the Bean-Livingston barrier \cite{Bean1964SurfaceBarrier} is overcome, making the Meissner state absolutely unstable to vortex nucleation resulting in substantial power dissipation \cite{gurevich2008dynamics}. Surface roughness introduces two effects that reduce the field limit of the metastable Meissner state: magnetic field enhancement and superheating field suppression. Each can introduce excess dissipation and, through thermal feedback, trigger thermal instability resulting in cavity quench \cite{gurevich2008dynamics,Xie2011QuenchSimulation}. Preventing nucleation of flux may assist in maintaining high peak accelerating fields in N-doped, low-temperature baked Nb and Nb$_3$Sn \cite{gurevich2008dynamics,pack2020vortex,Carlson2021AnalysisOfDissipationInNb3Sn,Lechner23TopographicEvolution,Lechner2024OxideDisScenarios,Lechner2025Nb3SnTopo}.

Previous investigations have used white light interferometry, atomic force microscopy and stylus profilometry to examine topographic differences due to electropolishing in either as-received or relatively rough mechanically ground Nb \cite{Xu2011Enhanced,Xu2012TopographicPSDStudy,tian2011novel}. In this work, we take a fundamentally different approach to examining surface roughness on polished Nb. Starting from a highly polished Nb surface, we investigate the defects introduced by electropolishing rather than the defects eliminated from the as-received surface. Such a strategy is required to understand the ultimate limits of any polishing process since large slowly undulating topography from the as-received surface can obscure the small, but sharp features that limit ultimate stability of the Meissner state. Using this method, we reveal the development of sharp intergranular sloped steps in electropolished Nb and calculate the associated magnetic field enhancement and superheating field suppression factors associated with such a geometry with the London theory \cite{london1935electromagnetic, Bean1964SurfaceBarrier}. Finally, the impact of surface roughness on impurity diffusion is investigated.

\section{\label{sec:Methods} Methods}

\subsection{Atomic Force Microscopy}
Atomic force microscopy (AFM) measurements were made in tapping mode using a Digital Instruments Nanoscope IV microscope on loan to Thomas Jefferson National Accelerator Facility from The College of William and Mary. The silicon AFM probe features a tip radius less than 10 nm. Topographic images were acquired over a 10 $\mu$m $\times$ 10 $\mu$m area consisting of 512 $\times$ 512 pixels resulting in a point spacing of $\sim$20 nm. The tip hosts a half-cone angle between 20° and 25° along the cantilever axis, 25° and 30° from the side and 10° at the apex.

\subsection{White Light Interferometry}
White light interferometry (WLI) measurements were conducted at the College of William \& Mary using a Profilm 3D optical profiler in phase shifting interferometry (PSI) mode, to achieve sub-nm vertical resolution. Measurements were made over a 2.0 mm $\times$ 1.7 mm area using a Nikon 10$\times$ objective with a lateral spatial resolution of 3.5 {\textmu}m.

\subsection{Sample Preparation}

Polycrystalline fine-grain Nb samples were cut by wire electro-discharge machining (EDM) to 1 inch diameter disks from Nb sheets ($RRR$=300), which were prepared following the requirements of the XFEL specification \cite{singer2015superconducting}. The typical grain size in this material is $\approx$ 50 $\mu$m, however significant variations in grain size can be found in the same material which can be linked to the rolling and recrystallization processes \cite{bieler:srf09-tuoaau03, grill:srf11-thpo059,Balachandran2021DirectEvidence}. The samples were then polished using BCP in a solution of HF (49\%), HNO$_3$ (70 \%), and H$_3$PO$_4$ (85\%) mixed in a 1:1:1 volume ratio to remove 50 {\textmu}m from the surface. Subsequently, the Nb samples were mechanically polished (MP) on one side to improve homogeneity and topography before EP. The bulk Nb samples were prepared using a Buehler Automet 300 with the Buehler commercial consumables following procedure outlined in the Supplementary Material. The existing polishing procedure produces an initial average roughness of below 5 nm over an area of 1690 $\mu$m $\times$ 1690 $\mu$m, based on the WLI method. In comparison, SRF cavities processed via a combination of centrifugal barrel polishing with EP would have a roughness of the order of 100 nm over an area of 270 $\mu$m $\times$ 270 $\mu$m \cite{prudnikava2018}. The surface roughness of the samples used in this work, prior to electropolishing, is significantly smoother than that typically encountered in SRF cavities. The optical imaging, scanning electron microscopy imaging, and electron backscatter diffraction (EBSD) orientation image maps (OIM) were performed on the surface to evaluate the material quality after the MP procedure, see the Supplementary Material.

\subsection{Electropolishing}
\begin{figure}
\centering
\includegraphics[width = 8.5 cm]{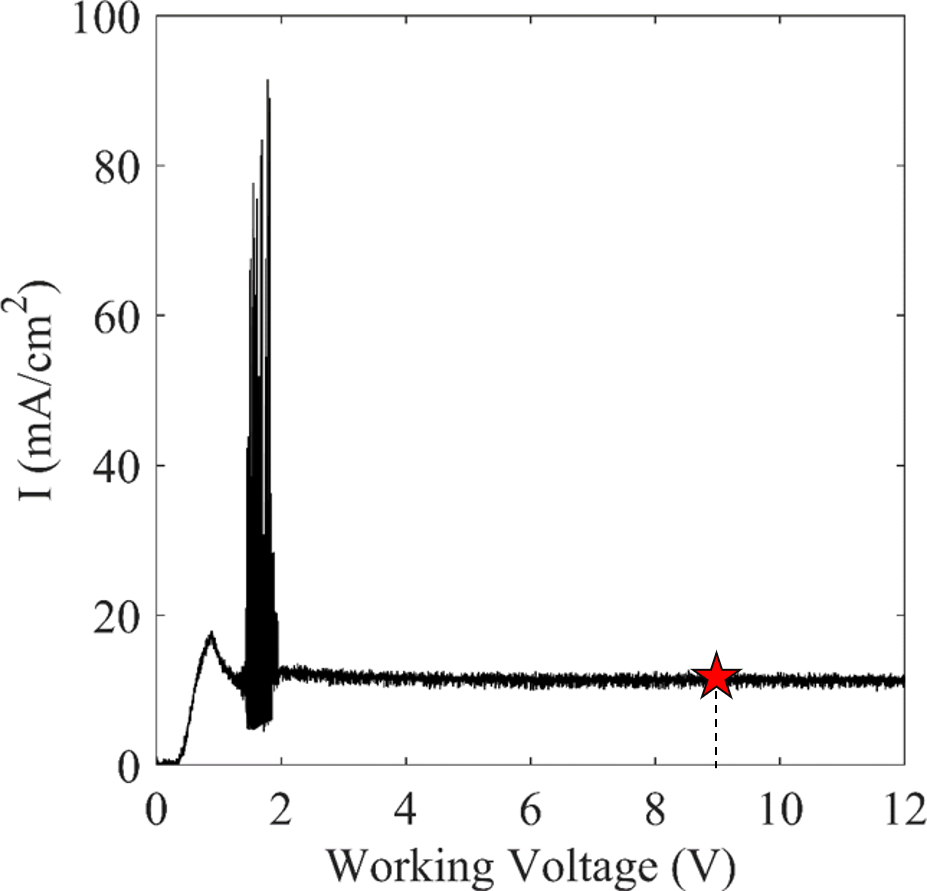}
\caption{Representative I-V characteristic of the electropolishing process. A red star represents the selected operating voltage. Non-equilibrium oxide growth and the diffusion-limited oxide dissolution result in the current oscillations near 2 V.}
\label{FigIV}
\end{figure}

Here and throughout, the term 'electropolishing' is used in the practical sense. In electrochemistry, electropolishing denotes the potential-driven anodic smoothing process whereas in practice, as applied SRF cavities, the process also proceeds with a parallel etching, as evidenced by rotating disk electrode measurements \cite{Tian2010EvaluationOfDiffusionCoefficient}. To investigate topographic defects introduced by electropolishing we electropolished the samples in a 1 to 9 by a volume mixture of HF(49\(\%\)) to H$_2$SO$_4$(98\(\%\)). Samples 1 inch in diameter and $\approx$3 mm thick were wrapped in PTFE tape, mounted in a sample holder, and immersed in the electropolishing electrolyte, allowing only the polished face to be exposed to the electrolyte. The electropolishing current was recorded using a digital multimeter and the material removed was determined via Faraday's laws of electrolysis. A coefficient of five electrons per Nb atom removed was used in the calculation, which was confirmed via weight loss measurements. Samples were electropolished at 13 °C and 9 V. The EP temperature is similar to that chosen for the cold EP process used in the LCLS-II cavities \cite{Posen2022HighGradientPerformanceCM,Furuta2019FermilabEPFacility,Crawford2017ExtremeDiffLimited}. The voltage selected is well within the "plateau" region of the I-V curve for our bench electropolishing geometry as shown in Fig. \ref{FigIV}. Electropolishing was performed in static electrolyte to ensure uniform removal across the surface. The temperature of the electrolyte was regulated via an aluminum heat exchanger cooled by a chiller.

For all electropolishing runs, the cell (EP electrolyte, Al cathode and Nb anode) was initially set up with the circuit open. The power supply voltage was first adjusted to the desired value, and then the anode lead was connected to close the circuit and begin polishing. At the end of a run, the process was terminated by switching off the power supply to reduce the applied voltage to zero, and then immediately disconnecting the anode lead.

We employed two complementary approaches to investigate the effects of electropolishing on surface roughness: $random$ $sampling$ and $sequential$ $electropolishing$. In the random sampling approach, samples from the same batch underwent varying amounts of electropolishing, enabling the study of diverse microstructures without tracking specific grains. In contrast, the sequential approach applied successive electropolishing steps to the same sample. After each electropolishing step, topography measurements were collected from the same set of grains, which controlled for microstructure.

\section{\label{sec:ResultsAndDiscussion} Results and Discussion}

\subsection{Surface Roughness Characterization}

\begin{figure}
\centering
\includegraphics[width=1\linewidth]{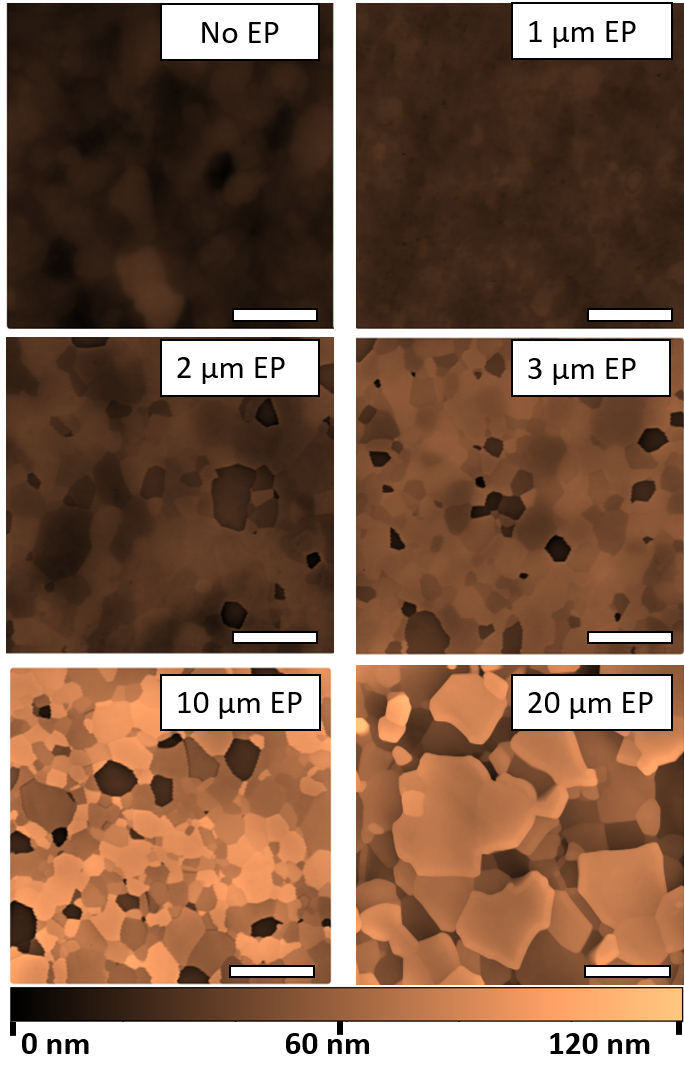}
\caption{WLI images of mechanically polished Nb samples subjected to increasing electropolishing removal. The WLI images show a development of topography due to the electropolishing process. Scale bars are 500 µm.}
\label{fig1}
\end{figure}

\begin{figure}
    \centering
    \includegraphics[width=1\linewidth]{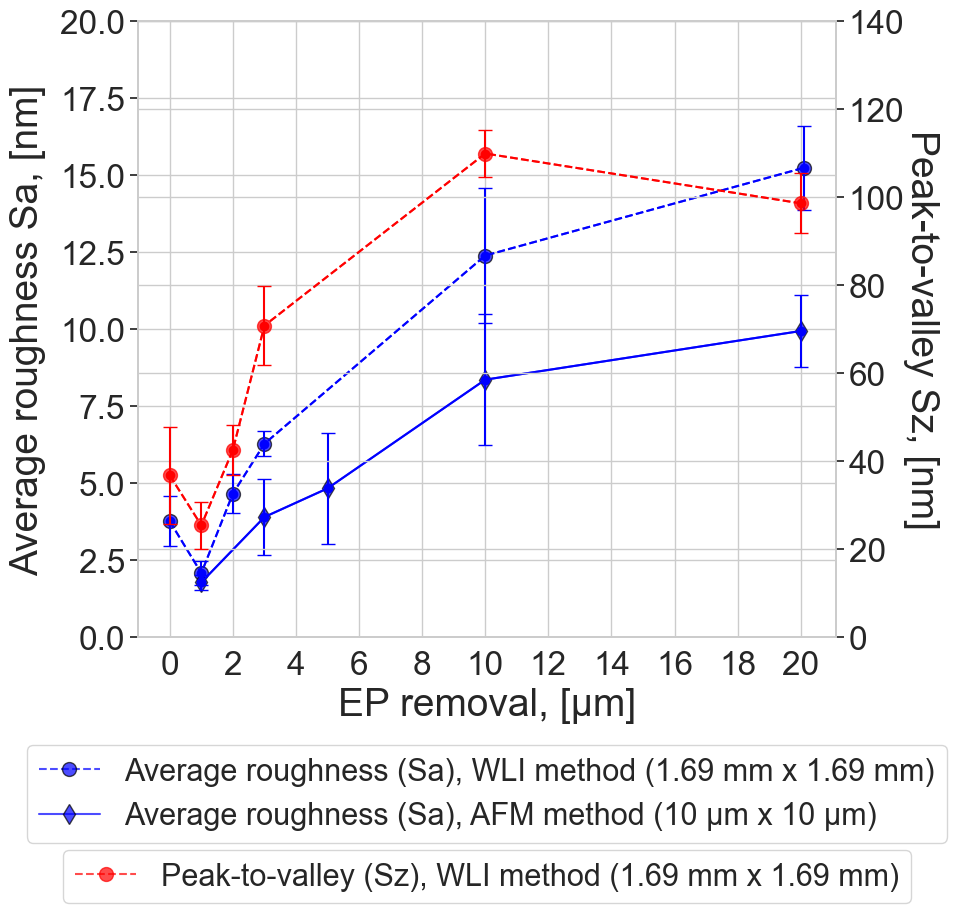}
    \caption{Progression of the average roughness ($S_a$) and peak-to-valley roughness parameter ($S_z$) with EP removal.}
    \label{fig14}
\end{figure}

To examine the development of topographic defects due to electropolishing, the mechanically polished surfaces were subjected to various amounts of electropolishing and investigated via WLI and AFM. Our WLI measurements of samples electropolished up to 20 $\mu$m reveal the growth of intergranular steps as shown in Fig. \ref{fig1}. The residual surface height variation after electropolishing is on the order of $\sim$1\% of the total removal depth, arising primarily from intergranular steps. Interestingly, although the samples are cut from the same batch, the sample subjected to 20 $\mu$m EP presents larger grains. This phenomenon can be attributed to rolling and recrystallization processes, which demonstrate the influence of the local microstructure \cite{bieler:srf09-tuoaau03, grill:srf11-thpo059, Balachandran2021DirectEvidence}. The average roughness ($S_a$) and peak-to-valley roughness ($S_z$) tend to grow with increasing EP material removal, as shown in Fig. \ref{fig14}.

It is important to note that $S_a$ and $S_z$ are significantly influenced by the size of the scan area \cite{tian2011novel, comparisonSaSz}. Generally, the measured roughness increases with a larger scan size, as illustrated by the increase in surface roughness measured by WLI compared to AFM in Fig. \ref{fig14}. Since no single instrument can probe roughness on the millimeter lateral scale with a few nanometer lateral resolution, it is often useful to unify different topographic measurements using a power spectral density (PSD) analysis \cite{Elson:95, Duparre:02, Jacobs_2017}. Although PSD is a widely used tool to characterize surface roughness, its utility in the context of SRF cavity performance is fundamentally limited. A detailed discussion of the limitations of PSD can be found in the Supplementary Material. PSD captures only the magnitude of spatial frequency components while discarding phase information. It lacks any direct connection to the performance-relevant effects of magnetic field enhancement or vortex nucleation and thus provides little guidance on which defects are the primary drivers of these phenomena. However, it can be a useful metric if the defects do not have a regular structure. The topographic defects described in this work are well defined. For these reasons, the PSD analysis is relegated to the Supplementary Material.

\subsection{Characterization of the Sloped Step Defect}
\begin{figure}
    \centering
    \includegraphics[width=1\linewidth]{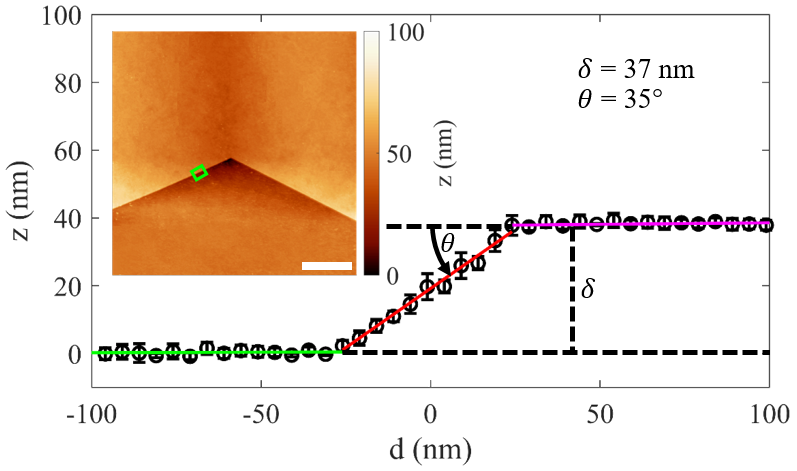}
    \caption{A representative profile across a grain boundary is shown. This profile is derived from the topography in the inset using the area enclosed in green and measures the intergranular step ($\delta$) and slope angle ($\theta$). The scale bar in the inset is 2 µm.}
    \label{fig2}
\end{figure}

A precise determination of the slope angle at the intergranular steps is not possible using WLI due to its insufficient lateral spatial resolution. We further investigated the topography of these steps using atomic force microscopy which revealed the existence of high slope angles. To quantify the slope angle, $\theta$, and the height of the intergranular step, $\delta$, a topographic profile across the grain boundary is examined. The profile is constructed by defining a line along a grain boundary and calculating the normal lateral displacement, $d$, between the line along the grain boundary and the surrounding pixels. The vertical displacement, $z(d)$, is binned in 5 nm increments from the grain boundary to a distance of 200 nm from the grain boundary. This analysis is shown in Fig. \ref{fig2} which reveals the existence of a high slope angle at the grain boundary. The observed slope angles are much larger than the ones previously reported \cite{knobloch1999high,Xu2011Enhanced} which is due to our higher spatial resolution and improved analysis. 

\begin{figure}
    \centering
    \includegraphics[width=1\linewidth]{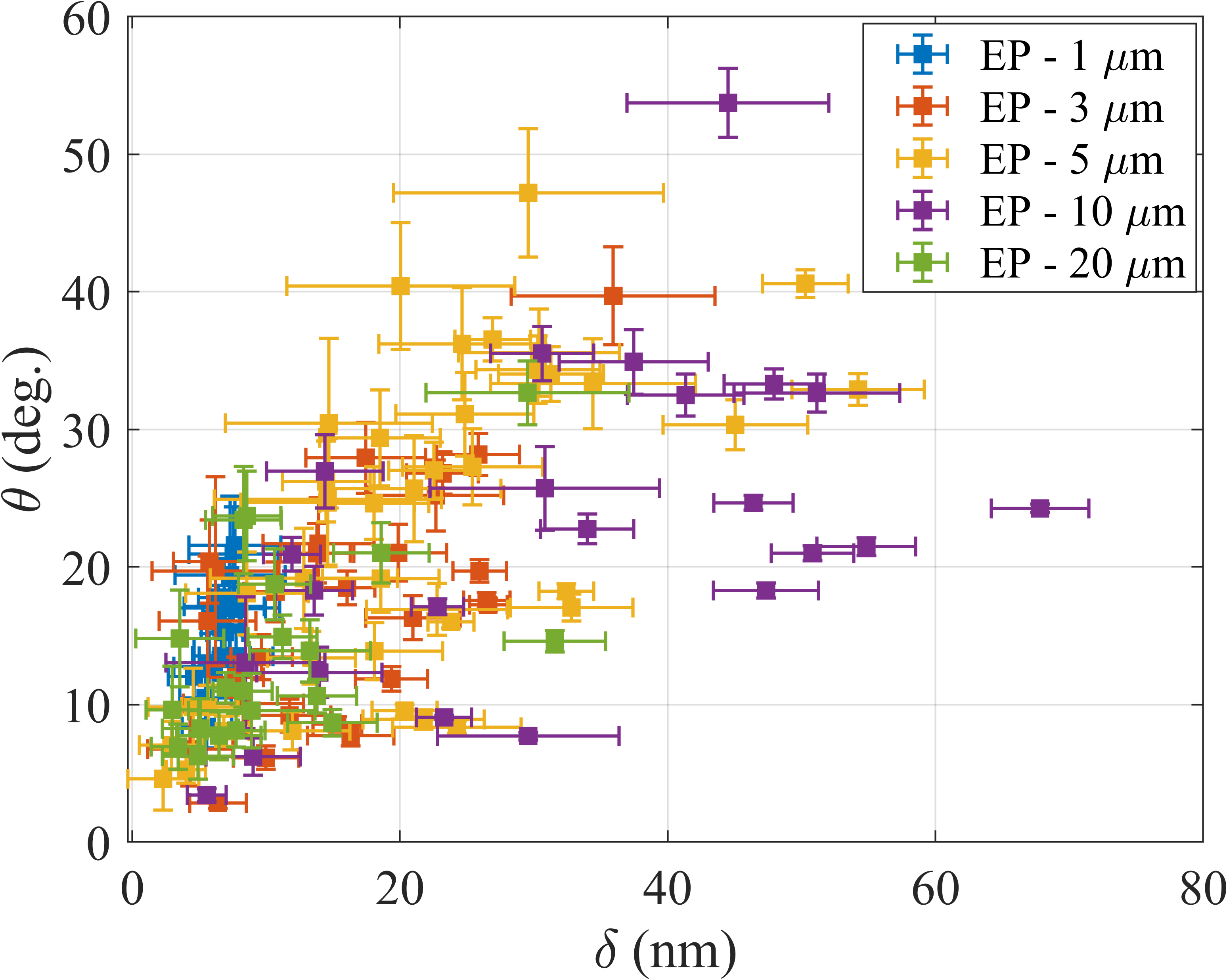}
    \caption{Evolution of a slope angle ($\theta$) and an intergranular step ($\delta$) upon EP using random sampling.}
    \label{fig3}
\end{figure}

The topographic evolution upon electropolishing was investigated using two different approaches. The first approach involved 10 random topographic measurements at grain boundary junctions for each EP removal. The measurements of the slope angle and intergranular step height using the random sampling approach are shown in Fig. \ref{fig3}. The intergranular steps left behind by the electropolishing process can be substantial, up to 70 nm, and the local slope angle can reach up to $\sim$50°. Qualitatively, with increasing EP removal the distribution of intergranular steps and slope angles appears to widen, yet at 20 $\mu$m of EP removal, the clustering is observed to be tighter. Based on the observation that the 20 $\mu$m EP removal sample presented a significantly different microstructure in the form of larger grains as shown in Fig. \ref{fig1}, it is possible that these distributions are primarily associated with microstructure. This apparent improvement may be the result of a reduction in the nearest-neighbor grain misorientation. No clear dependence of slope angle and step heights on electropolishing removal is observed when randomly sampled.

\begin{figure}
    \centering
    \includegraphics[width=1\linewidth]{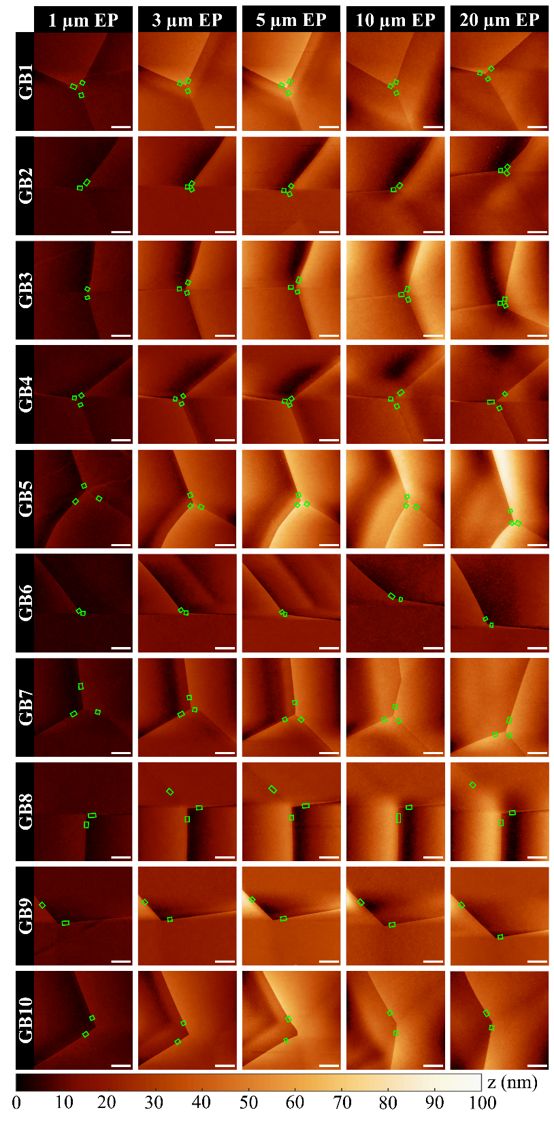}
    \caption{AFM topographies showing the evolution of grain boundaries (GB) upon electropolishing. A green rectangular outline indicates the region used for the calculation of an intergranular step ($\delta$) and a slope angle ($\theta$). The scale bars are 2 µm.}
    \label{fig4}
\end{figure}

\begin{figure}
    \centering
    \includegraphics[width=1\linewidth]{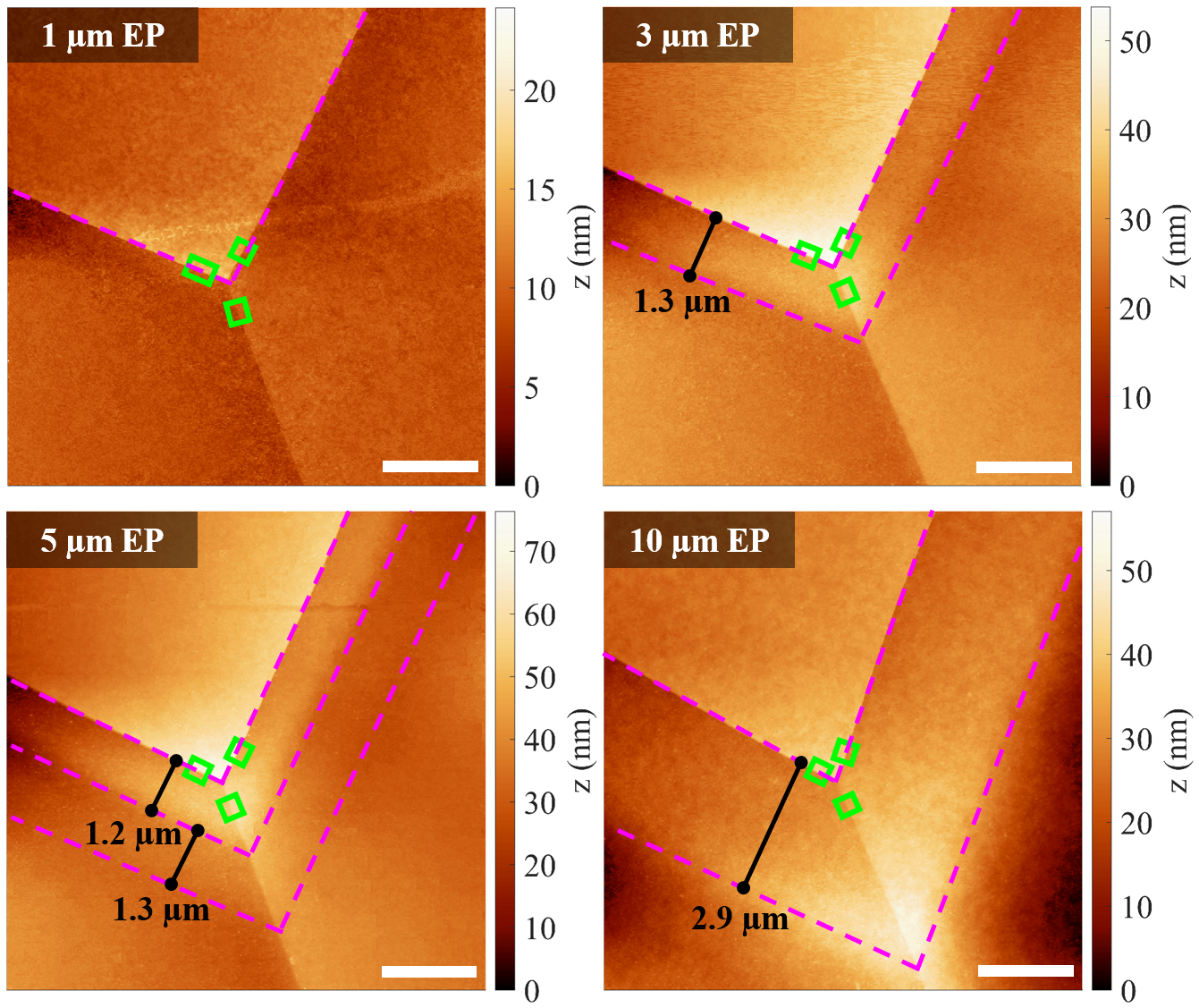}
    \caption{AFM topographies showing the evolution of grain boundary (GB1) upon electropolishing. A green rectangular outline around the grain boundary indicates the area chosen to construct a line profile used for a calculation of an intergranular step ($\delta$) and a slope angle ($\theta$). The dashed magenta lines show the evolution of the grain boundaries. The scale bars are 2 µm.}
    \label{fig5}
\end{figure}

The second approach involved sequential EP of samples and examining the same grains after each EP cycle to control for the effects of microstructure apparent when sampling randomly. Grains were identified by correlating the optical image with the positions of the surrounding grains and grain boundaries. Representative sequentially electropolished grain boundaries are shown in Fig. \ref{fig4}. At 1 $\mu$m removal, Fig. \ref{fig4} shows the development of intergranular steps. Upon additional electropolishing, these steps are smoothed, but remnants of their existence persist in the form of a smoothed line parallel to the new location of the grain boundary. Those types of remnants are clearly observable in GB 1, 5, 6 and 10. Examining the progression of GB 1 in more detail, the stripes surrounding the grain boundary lines have widths related to the EP removal increment, a $\sim$0.6 $\mu$m shift per $\mu$m removed, as shown in the 3, 5 and 10 $\mu$m removal images in Fig. \ref{fig5}. The shift in the apparent grain boundary location is due to the fact that the grain boundary extending toward the bulk is not oriented normal to the surface. Upon deep enough electropolishing, the appearance of the grain boundary can change as highlighted by the 10 and 20 $\mu$m AFM topographies of GB 1 and 7 in Fig. \ref{fig4}. With each electropolishing, the previous intergranular step is smoothed, demonstrating diffusion-limited electropolishing, but a new step always appears. Since the displacement of the step from its previous location is correlated with the EP increment, as shown in Fig. \ref{fig5}, this indicates that the step is only formed at the end of the electropolish. Its development may be due to orientation dependent oxide growth or dissolution rates. The intergranular step height may be exacerbated via the parallel etching process observed in rotating disk electrode experiments \cite{Tian2010EvaluationOfDiffusionCoefficient}. The parallel etching process more likely explains the development of grain specific heights observed in the WLI images. Etching from the post-EP, EP electrolyte exposure, common in electropolishing processes for SRF cavities \cite{Crawford2017ExtremeDiffLimited,Geng2011StandardHighGradProcedures,Song2021_650MHzDevelopment}, may also play a role. The distribution of slope angles and step heights in the sequential EP experiments are shown in Fig. \ref{fig6}. Again, no obvious trend of electropolishing removal is observed on the distribution of slope angles and step heights. Based on the tight clustering observed in Fig. \ref{fig6} it is clear that the intergranular step is primarily controlled by the relative grain orientation rather than the EP depth. Weld seams are also known to exhibit substantial roughness \cite{Wenskat2017OpticalRFLimitationsXFEL}, and their process-dependent microstructure formed during welding may aggravate the formation of intergranular steps upon electropolishing.

\begin{figure}
    \centering
    \includegraphics[width=1\linewidth]{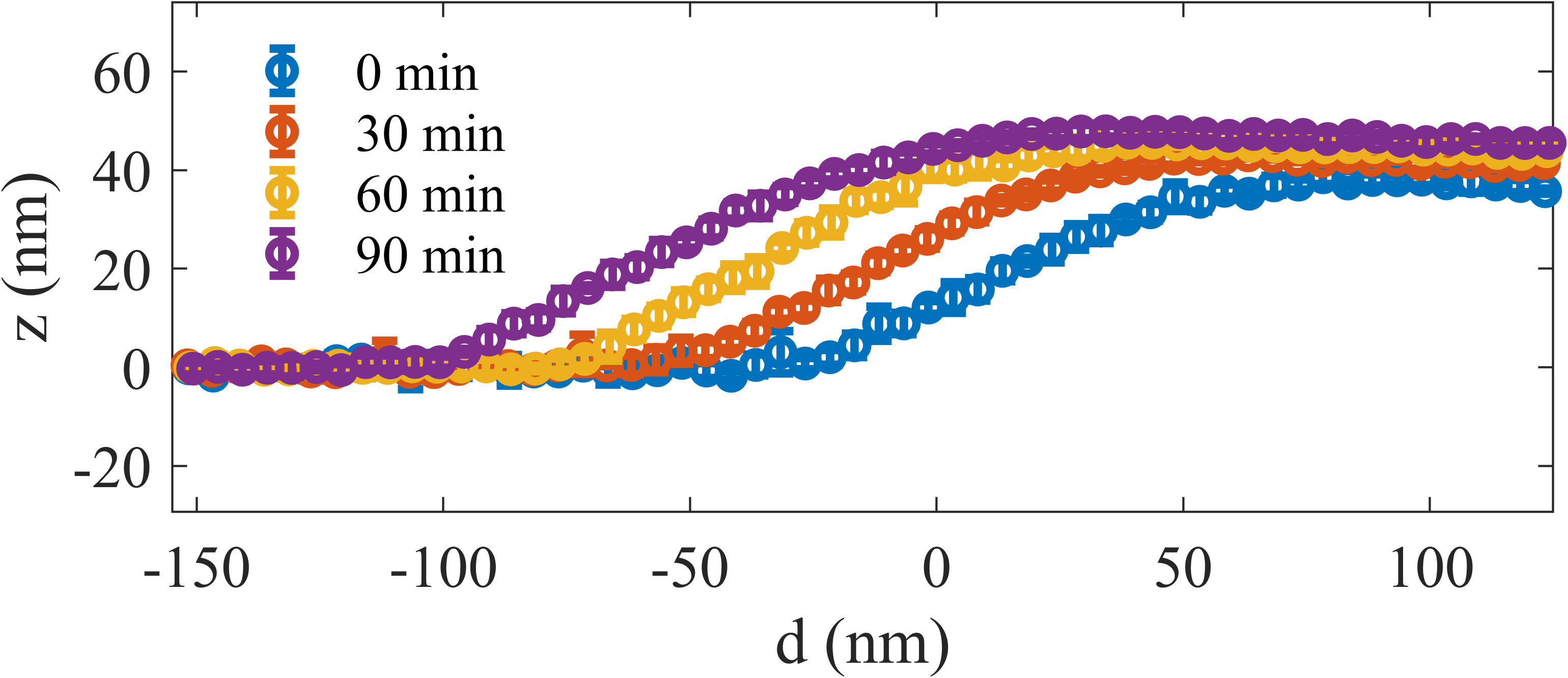}
    \caption{Evolution of intergranular step heights ($\delta$) upon different exposure times to the EP electrolyte. The line profiles are taken across the same grain boundary upon different exposure times (0, 30, 60, and 90 minutes). Some etching does occur which correlated with the exposure time. At this grain boundary the step height grew at a rate of 6 $\pm$ 2 nm/h. The temperature of the electrolyte is 13 °C, consistent with the temperature used during electropolishing.}
    \label{fig16}
\end{figure}

\begin{figure}
    \centering
    \includegraphics[width=1\linewidth]{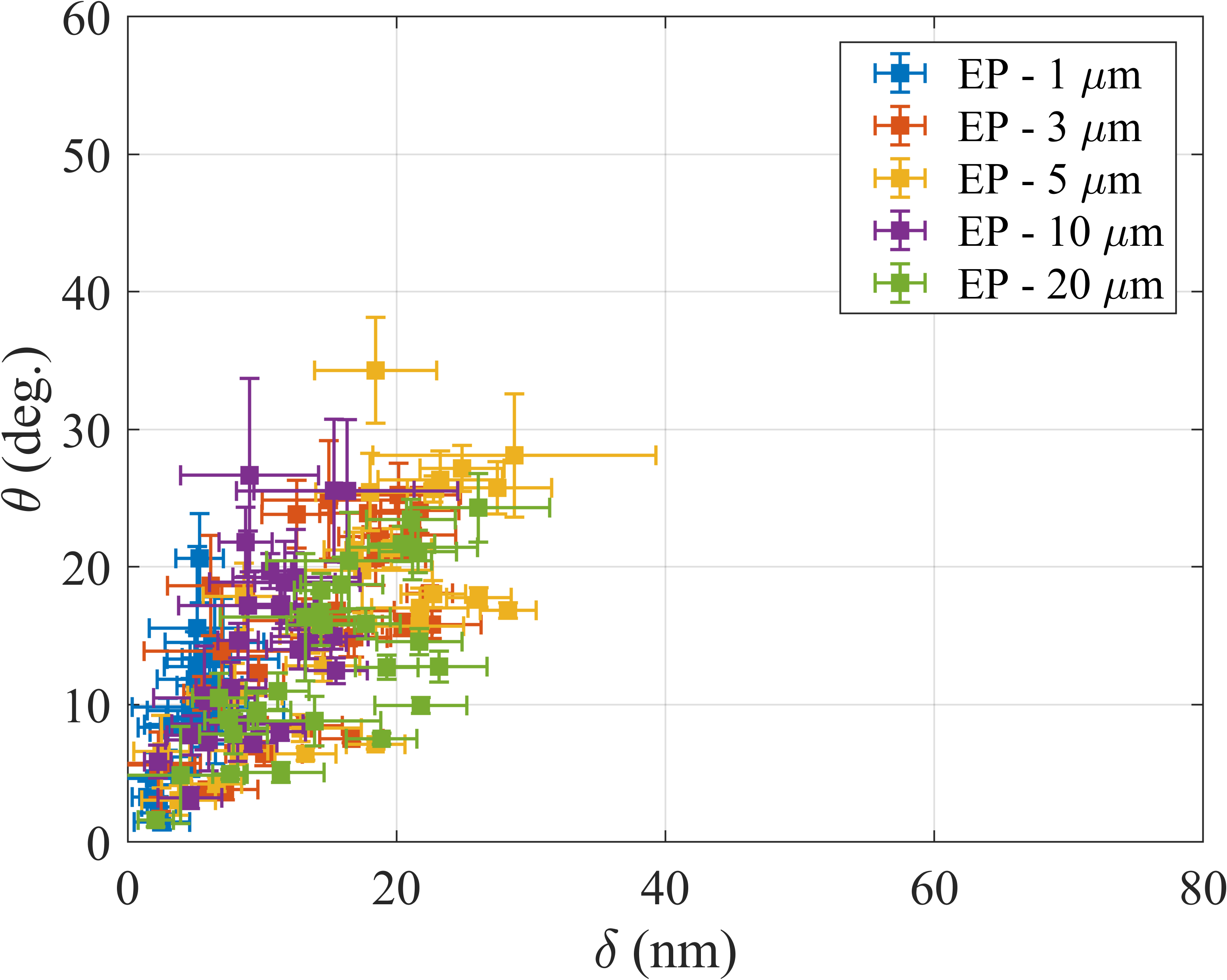}
    \caption{Evolution of a slope angle ($\theta$) and an intergranular step ($\delta$) upon sequential EP.}
    \label{fig6}
\end{figure}

We evaluated the effect of etching when the sample is simply exposed to the EP electrolyte (no applied voltage/electrodes not grounded) for three different durations: 30 minutes, 60 minutes, and 90 minutes. As shown in Fig. \ref{fig16}, the results indicate a small change in the intergranular step compared to the intergranular step resulting directly after EP, indicating that the electropolish process itself is the primary cause of the intergranular step and slope angle. The rate of increase in step height at this grain boundary junction was observed to be 6 $\pm$ 2 nm/h. Based on the grain orientation dependence observed in the aforementioned measurements, the step height growth rate is likely different at other grain boundaries.

These geometrical defects can introduce magnetic field enhancement, superheating field suppression and modify the impurity diffusion process. The calculated magnetic field enhancement, suppression factors associated with these topographic defects as well as the impact of topography on impurity diffusion are investigated in the next sub-sections. 

\subsection{\label{subsec:SFS}Superheating Field Suppression at Grain Boundary Steps}
\begin{figure}
\includegraphics[width=8.5cm]{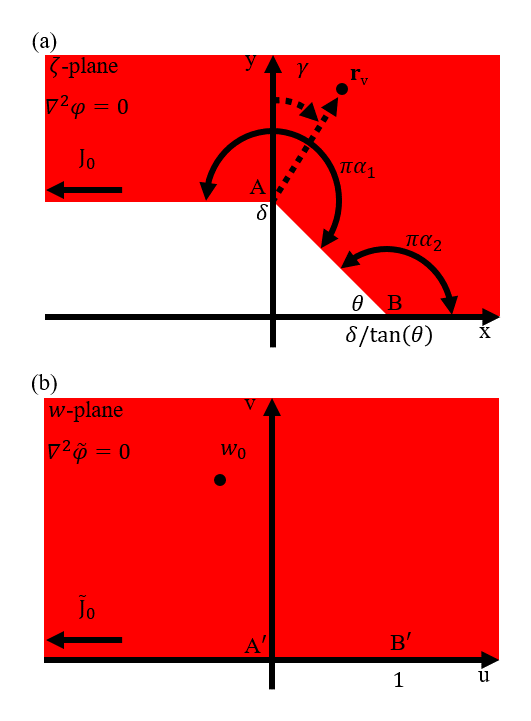}
\caption{(a) Real space representation of the grain boundary geometry and (b) the transformed geometry on the complex plane.}
\label{ModelGeometry}
\end{figure}

Surface roughness, like the intergranular sloped steps found in electropolished Nb, can cause a suppression of the superheating field. Theoretically, this defect can be roughly modeled by two parallel planes separated by a distance ($\delta$) and connected by a plane, at angle ($\theta$). We introduce a coordinate system with its origin placed directly below the large interior-angle corner of the step at point A in Fig. \ref{ModelGeometry} (a). The x-axis runs parallel to the surface of the grains, while the y-axis is normal to the surface. In this coordinate system, point A is located at $(0,\delta)$. The sloped surface descends linearly from point A to point B at $(\delta/\tan(\theta),0)$, where it meets the lower plane corresponding to the surface of the adjacent grain.

For this scenario, we consider an external magnetic field aligned along the z-axis and apply the London theory where the defect size is small compared to the penetration depth. A suppression of the superheating field arises as a result of current crowding around the defect, which enhances the force pushing a vortex into the superconductor, and a geometrical degradation of the surface barrier. For a vortex to penetrate the superconductor, the force from the surface barrier driving a vortex out of the material, $\mathbf{F}_S$, must be exceeded by the force driving the vortex into it, $\mathbf{F}_M$ \cite{buzdin1998ElectromagneticPinningOfVortOnDefects,kubo2015field}. The force of the surface barrier acting on the penetrating vortex and the Lorentz force from the screening current driving the vortex into the surface are given by

\begin{equation}
\mathbf{F}_S=\mathbf{J}_I \times \phi_0 \hat{\mathbf{z}},
\label{eqForceFromSurface}
\end{equation}

\begin{equation}
\mathbf{F}_M=\mathbf{J}_M \times \phi_0 \hat{\mathbf{z}}=\frac{\Gamma(\mathbf{r})B_a}{\mu_0\lambda} \mathbf{\hat{J}}_M \times \phi_0 \hat{\mathbf{z}},
\label{eqForceFromMagneticField}
\end{equation}
where $\mathbf{J}_M$ is the current density induced by the external magnetic field, $\mathbf{J}_I$ is the current density of the image vortex required to satisfy the boundary condition $(\mathbf{J}_{I}+\mathbf{J}_{V})\cdot \hat{\mathbf{n}}=0$, and $\mathbf{J}_V$ is the current density of the nucleated vortex.  $\phi_0$ is the magnetic flux quantum, $\Gamma\mathbf{(r)}$ is the local current density enhancement factor, $B_a$ is the applied magnetic field, $\mu_0$ is the permeability of free space and $\lambda$ is the London penetration depth. The current density enhancement factor accounts for the increase in current density near the defect relative to the uniform current density far from the geometrical defect, $J_M(\infty,y)=\frac{B_a}{\mu_0\lambda}$. For a perfectly flat surface, the superheating field in the London theory is \cite{gurevich2008dynamics}
\begin{equation}
    B_{v}=\frac{\phi_0}{4\pi\lambda\xi},
\label{eqLondonSuperheatingField}
\end{equation}
where the $\xi$ is the coherence length. At the geometrically suppressed superheating field, $B_a=B^*_v$. A superheating field suppression factor $\eta$, defined by $B^*_v=\eta B_{v}$, based on the topographic defect described here and on the condition that the surface barrier at a distance of one $\xi$ from the surface is overcome by the force from the external magnetic field, quantified by $|\mathbf{F_S}|=|\mathbf{F_M}\cdot\mathbf{\hat{F}_S}|$, is given by 
\begin{equation}
\begin{multlined}
\eta(\theta,\delta,\xi)= \\
    \min_\gamma \left( \frac{|\mathbf{J}_I(\mathbf{r}_v) \times \phi_0 \hat{\mathbf{z}}|}{|B_{v}\frac{\Gamma(\mathbf{r}_v)}{\mu_0\lambda} (\mathbf{\hat{J}_M(\mathbf{r}_v)} \times \phi_0 \hat{\mathbf{z}})\cdot(\mathbf{\hat{J}}_I(\mathbf{r}_v) \times \hat{\mathbf{z}})|}\right).
\end{multlined}
\label{eqSuperheatingFieldSupressionFactor}
\end{equation} 
Here, $\gamma$ is defined as the angle measured from the $y$-axis at point A that parameterizes the possible vortex nucleation positions, $\mathbf{r}_v=(\xi\sin\gamma,\delta+\xi\cos\gamma)$. These points lie one coherence length $\xi$ from the surface near the corner. The superheating field corresponds to the easiest direction for vortex entry, which is obtained by minimizing $\eta$ with respect to $\gamma$.

Since $\nabla \cdot \mathbf{J} = 0$ and under the assumption $\nabla \times \mathbf{J} \approx 0$ conformal mapping can be employed to determine the complex potential for the screening current density enhancement at the geometrical defect. Similarly, the current density of the image vortex, $\mathbf{J}_I$, needed to preserve the boundary condition $(\mathbf{J}_{I}+\mathbf{J}_{V}) \cdot \mathbf{\hat{n}}=0$, which accounts for the surface barrier, is calculated via conformal mapping \cite{buzdin1998ElectromagneticPinningOfVortOnDefects,kubo2015field}. The current densities are calculated via standard conformal mapping techniques \cite{zill2009ComplexAnalysis}. A detailed explanation of this type of approach can be found in Ref. \cite{kubo2015field}. The transformation from the upper-half plane, shown in Fig. \ref{ModelGeometry} (b), to the sloped-step geometry is given by the Schwarz-Christoffel mapping

\begin{equation}
\zeta=F(w)=K_1\int_{0}^{w}f(w')dw'+K_2,
\label{SchwarzChristoffel}
\end{equation}
where for the sloped-step geometry $f(w)=w^{\alpha_1-1}(w-1)^{\alpha_2-1}$. $K_1$ and $K_2$ are calculated from the following relations
\begin{equation}
i\delta=K_1\int_{0}^{0}f(w')dw'+K_2=K_2,
\label{eqK2}
\end{equation}
\begin{equation}
\frac{\delta}{\tan(\theta)}=K_1\int_{0}^{1}f(w')dw'+K_2.
\label{eqK1}
\end{equation}
The real part of the complex potential yields the current density scalar potential due to the external magnetic field which is given by 
\begin{equation}
\phi_M(x,y)=\phi_M(w)|_{w=F^{-1}(x,y)}= \text{Re}(K_1J_0w)|_{w=F^{-1}}(x,y).
\label{eqJSP}
\end{equation}

\begin{figure}
\includegraphics[width=8.5cm]{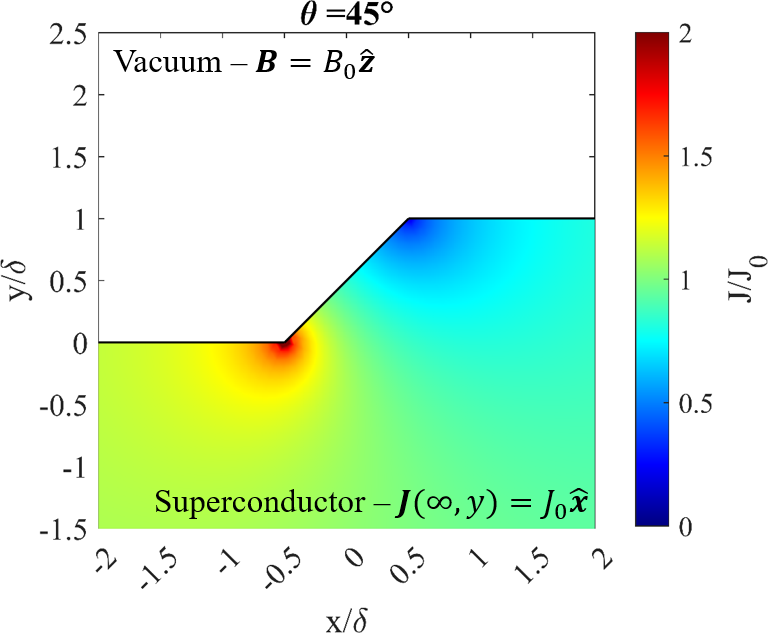}
\caption{Representative current density modification due to the sloped-step topographic defect. In this model, the external magnetic field is aligned along the z-axis inducing a screening current along the x-axis. Current crowding occurs at the region surrounding the larger internal angle.}
\label{FigCDEFExample}
\end{figure}

An increase in the current density is observed at the bottom of the sloped step as shown in Fig. \ref{FigCDEFExample}. The current density scalar potential due to the vortex and image vortex is given by

\begin{equation}
\phi_{V+I}(x,y)=\text{Re}\left( \frac{i\phi_0}{2\pi\mu_0\lambda^2}\ln\left(\frac{w-w_0}{w-w^*_0}\right)\right)\biggr|_{w=F^{-1}(x,y)}.
\label{eqJSPVAV}
\end{equation}
where $J_0$ is the uniform current density far from the defect, $w_0=F^{-1}(\mathbf{r}_v)$ is the cutoff position of the London theory in $w$-space and $w^*_0$ is the complex conjugate of $w_0$ which is the position of the image vortex that preserves the boundary condition of $(\mathbf{J}_{I}+\mathbf{J}_{V})\cdot\hat{\mathbf{n}}=0$ at the superconductor-vacuum interface. The current density scalar potential of the image vortex used to calculate the force from the surface is 
\begin{equation}
\phi_{I}(x,y)=\text{Re}\left( \frac{-i\phi_0}{2\pi\mu_0\lambda^2}{\ln(w-w^*_0)}\right)\biggr|_{w=F^{-1}(x,y)}.
\label{eqJSPAV}
\end{equation}
Eq. \ref{SchwarzChristoffel} - \ref{eqJSPAV} are similar to the expressions found in Ref. \cite{kubo2015field} only differing in the form of $f(w)$. Finally, to calculate Eq. \ref{eqSuperheatingFieldSupressionFactor}, the real-space current densities are $\mathbf{J}_I=-\nabla\phi_I$ and $\mathbf{J}_M=-\nabla\phi_M$.

\begin{figure*}
\includegraphics[width=1.0\textwidth]{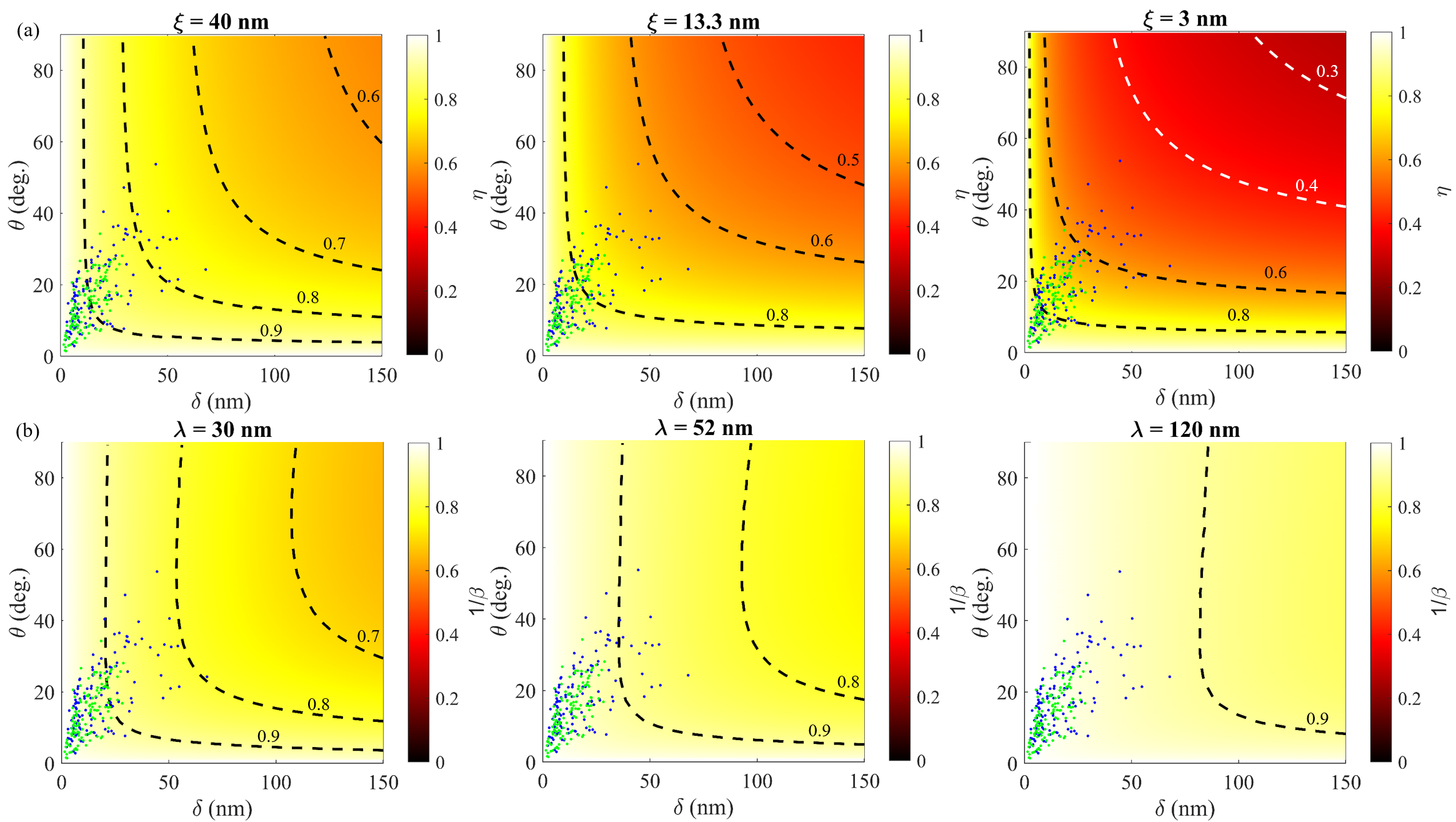}
\caption{(a) From left to right, $\eta(\theta,\delta)$  for $\xi =$ 40, 13.3, 3 nm representing clean Nb, alloyed Nb and next generation materials. (b) From left to right, $1/\beta(\theta,\delta)$  for $\lambda =$ 30, 52, 120 nm representing clean Nb, alloyed Nb and next generation materials. The blue and green points are the measurements from Fig. \ref{fig3} and Fig. \ref{fig6} respectively.}
\label{GBetaLandscape}
\end{figure*}

In the framework of this model, the superheating field suppression factor only depends on the grain boundary slope angle, intergranular step height and coherence length. $\eta(\theta,\delta,\xi)$ is mapped out for different values of $\xi$ representative of clean, ideally alloyed $(l\approx\xi_0/2)$ \cite{kubo2022effects} and next generation materials \cite{valente2016superconducting} as shown in Fig. \ref{GBetaLandscape} (a). The measured slope angles and intergranular step heights are also plotted in Fig. \ref{GBetaLandscape} (a). The superheating field is preserved as long as the slope angles and step heights are small. However, the observed slope angles and step heights shown in Fig. \ref{GBetaLandscape} (a) can contribute to a substantial degradation of the achievable peak field. As the coherence length shrinks the superconductor becomes more even sensitive to superheating field suppression. It is possible this mechanism is in part responsible for the degradation of peak magnetic fields in heavily N-doped cavities \cite{Gonnella2016ImprovedNDopingProtocols}, but the topographic defects revealed upon electropolishing N doped Nb, which also depend on the N-doping process or electropolishing process applied \cite{Lechner23TopographicEvolution, Chouhan2025MitigationOfPitting}, make it difficult to disentangle the effects.

\subsection{Magnetic Field Enhancement at Grain Boundary Steps}

\begin{figure}
\includegraphics[width=8.5cm]{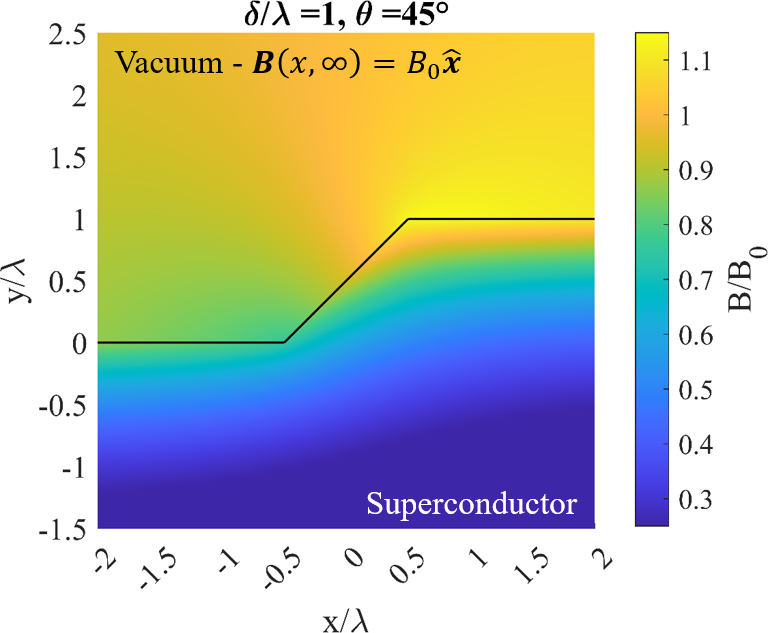}
\caption{Representative example of MFE due to a sloped step in the London model. In this model, the external magnetic field is aligned along the x-axis. The screening current flows in the $-\mathbf{\hat{z}}$ direction.}
\label{MFEExample}
\end{figure}

When topographic defects have a small radius of curvature comparable to or smaller than the length scale of the London penetration depth, the perfect electrical conductor model overestimates magnetic field enhancement. A better estimate for the magnetic field enhancement occurring at sharp edges can be obtained from the London model. The London model requires a two-domain solution. Above the superconductor, in vacuum, the magnetic vector potential, $\mathbf{A}$, is governed by
\begin{equation}
\nabla^2 \mathbf{A} = 0
\label{MagneticVectorPotentialAbove}
\end{equation}
and within the superconductor, the magnetic vector potential is governed by
\begin{equation}
\nabla^2 \mathbf{A} = \frac{1}{\lambda^2}\mathbf{A}.
\label{MagneticVectorPotentialWithin}
\end{equation}
In this two-dimensional geometry where the external magnetic field is along the x-axis, only the z-component of the vector potential, $A_z(x,y)$ is relevant. Consequently, the governing equations reduce to a scalar form for $A_z$ which automatically satisfies the Coulomb gauge, $\nabla \cdot \mathbf{A}=0$. The domains are complemented with the following boundary conditions. Far above ($y=150\lambda+\delta$ for this simulation) the superconductor the magnetic vector potential is $\mathbf{A}=B_0y\mathbf{\hat{z}}$. The magnetic field in vacuum far from the surface is $\mathbf{B}=B_0\mathbf{\hat{x}}$, where $B_0$ is the magnitude of the magnetic field far from the surface. The magnetic field at the surface was calculated via $ \mathbf{B}=\nabla \times \mathbf{A}$. The magnetic vector potential is continuous across the superconductor-vacuum interface. Deep ($y=-5\lambda$ for this simulation) in the superconductor $\mathbf{A}=\mathbf{0}$. At the boundary, $x=x_{min}$ and $x=x_{max}$ in and out of the superconductor, $\nabla A_z \cdot \mathbf{\hat{x}}=0$. Since Eq. \ref{MagneticVectorPotentialWithin} can be made dimensionless with the following transformations $\mathbf{r}/\lambda \rightarrow \Tilde{\mathbf{r}}$, $\lambda\nabla \rightarrow \Tilde{\nabla}$ and $\mathbf{A} \rightarrow B_0\lambda\Tilde{\mathbf{A}}$, magnetic field enhancement (MFE) in the London model depends on the ratio $\delta/\lambda$ and the slope angle. An example of the magnetic field enhancement at grain boundary steps is shown in Fig. \ref{MFEExample}. The magnetic field enhancement factor, $\beta$, of intergranular step heights of 1 to 150 nm and slopes from 1 to 90 degrees were simulated with different values of the penetration depth representing clean limit Nb, estimated at 30 nm \cite{McFadden2023DepthResolvedMeissnerScreeningProfInNb,Prozorov2022NbCleanLimit}, "dirty" Nb and next generation materials as shown in Fig. \ref{GBetaLandscape} (b). A natural consequence of the dependence of MFE on $\delta/\lambda$ is that as the mean free path is reduced via impurity alloying techniques, the superconductor's penetration depth increases and the material becomes \emph{less sensitive} to geometric magnetic field enhancement since the ratio $\delta/\lambda$ reduces. From the perspective of the effect of MFE on the superconducting state, the alloyed surface is effectively smoother compared to the clean limit. This is in stark contrast to the superheating field suppression factors, which become more severe with alloying as shown in Fig. \ref{GBetaLandscape}. The dependence of MFE on $\delta/\lambda$ can be found in the Supplementary Material.

We suspect that the superheating field suppression factor (SFS) model underestimates the suppression factor in the clean limit. Based on the similarity of the clean limit SFS and MFE in Fig. \ref{GBetaLandscape} (a) and (b), magnetic field enhancement may dominate the high field losses near the clean limit. There may be a cross-over between MFE and SFS induced losses as the material falls deeper into the type-II regime. Again, it is interesting to note that the MFE factors become less severe as the superconductor's electron mean free path shrinks. This mechanism may be behind the small increase in accelerating gradient before high field Q-slope develops in O-alloyed mid-T baked \cite{posen2020ultralow,Ito2021FurnaceBaking,lechner2021rf} Nb cavities  \cite{steder2024furtherimprovement} and a slightly enhanced vortex penetration field in BCP'd Nb subjected to mid-T bake \cite{Thoeng2024MeissnerScreeningBreakdown}. It may also play some role in low-temperature baking.

\subsection{Impurity Migration at Topographic Defects: Consequences for Shallow Impurity Profiles}

\begin{figure}
\includegraphics[width=8.5cm]{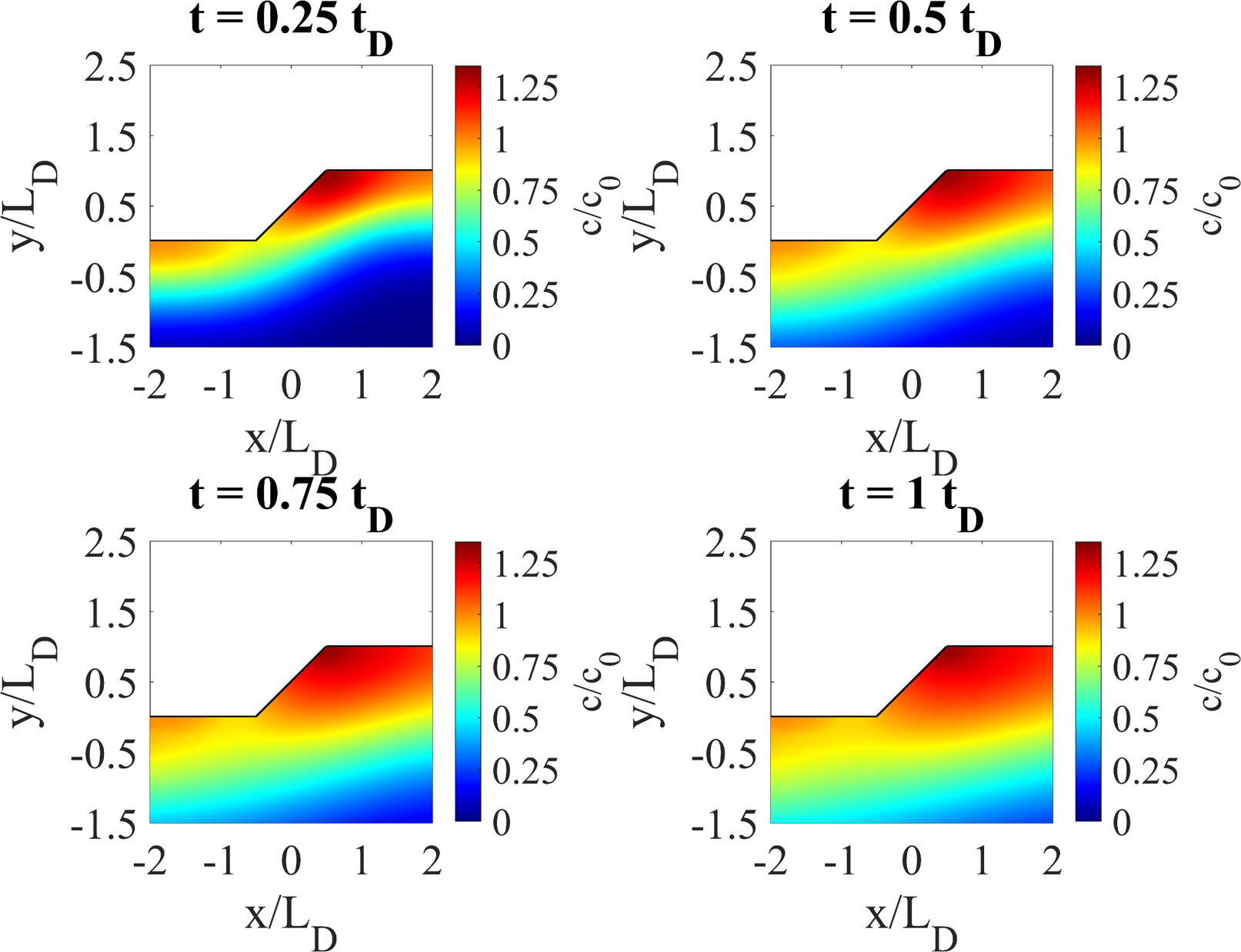}
\caption{Snapshots of impurity diffusion around the sloped step geometry in the finite impurity source model.}
\label{DiffusionSnapshots}
\end{figure}

Migration of impurities near topographic defects hosting large slope angles can cause a local depression in the concentration of impurities due to the expansion of impurities into regions with large interior angles. To simulate this effect we utilize Fick's second law of diffusion,
\begin{equation}
\frac{\partial c}{\partial t}  = D\nabla^2c,
\label{diffusionEquation}
\end{equation}
where $c$ is the concentration of the impurities, and $D$ is the diffusion coefficient on the sloped-step geometry. Equation \ref{diffusionEquation}, made dimensionless with the following transformations $\mathbf{r}/L_D \rightarrow \Tilde{\mathbf{r}}$, $L_D\nabla \rightarrow \Tilde{\nabla}$, $t/t_D \rightarrow \Tilde{t}$ and $c/c_0 \rightarrow \Tilde{{c}}$, predicts the diffusion of impurities that depends on the slope angle and the normalized height of the intergranular step, $\delta/L_D$. Here, the diffusion length is defined by $L_D=\sqrt{Dt_D}$ and $t_D$ is the diffusion time. The system is complemented by the following boundary conditions $c(x, -\infty)=0$ and, far from the defect, $\nabla c\cdot\mathbf{\hat{x}}=0$. The surface's boundary condition depends on the impurity diffusion scenario considered. Three impurity diffusion scenarios have been simulated, a finite impurity source on the surface ($\nabla c(x,h(x))\cdot\mathbf{\hat{n}}=0$ and $c(x,h(x),t=0)=\nu_0\delta(y-h(x))$, where $h(x)$ is the surface contour), a constant flux of impurities from the surface ($\nabla c(x,h(x))\cdot\mathbf{\hat{n}}=\phi$), and a constant impurity concentration at the surface ($c(x,h(x))=c_0$). Here we present the results of the finite impurity source model \cite{Ciovati2006ImprovedODiffusion}. The results of the other scenarios, which are qualitatively similar, can be found in the Supplementary Material. Snapshots of the impurity diffusion process near the topographic defect are presented in Fig. \ref{DiffusionSnapshots}. It is observed that the impurity concentration is reduced near the bottom of the sloped step as a consequence of the larger interior angle that provides a greater volume for the impurities to expand into. Inversely, the top of the topographic defect can host an enhanced impurity concentration compared to the regions far from the defect as a consequence of diffusion into a smaller interior angle.

As first pointed out by Kubo \cite{kubo2015field,KuboMultilayerReview2016}, low-temperature baked Nb \cite{Ciovati2004EffectOfLTbaking}, due to the shallow migration of oxygen during the treatment, can be roughly considered as a dirty superconductor/superconductor bilayer theoretically capable of extending the metastability of the Meissner state to higher fields. Many works have refined theoretical insights on the superconductor/superconductor bilayer and thin impurity profile systems capable of extending the field limit of the Meissner state \cite{Sauls2019EffectOfInhomogeneous,Kubo2021SuperheatingFieldsOfSemiinfinite,Pathirana2023SuperheatingFieldNanostructured}. The "dirtiness" of the top layer is one of the important parameters in this model. Neglecting the deleterious effects of impurities \cite{Desorbo1963EffectOfDissolvedGasesInNb} on the superconductivity of Nb, theoretically the dirtier the top layer, the higher the field limit \cite{KuboMultilayerReview2016}.
In light of the superheating field suppression model described in Section \ref{subsec:SFS}, vortex nucleation is expected to occur directly below the bottom of the sloped step, which coincides with where the impurity content is suppressed. 

\begin{figure}
\includegraphics[width=8.5cm]{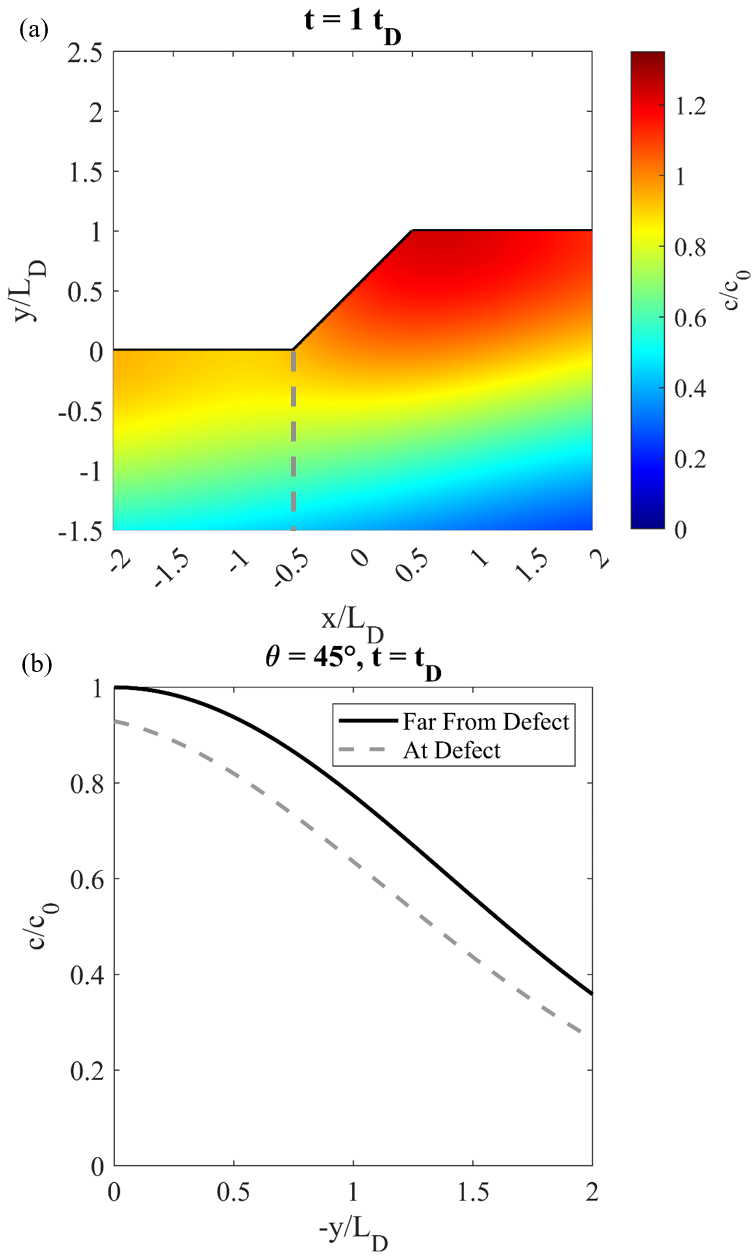}
\caption{(a) Solution of Eq. \ref{diffusionEquation} the gray line extending toward the bulk indicates the location of the line profile plotted in (b). (b) Comparison between impurity profiles at the defect bottom and far from the defect.}
\label{DiffusionComparison}
\end{figure}

\begin{figure}
\includegraphics[width=8.5cm]{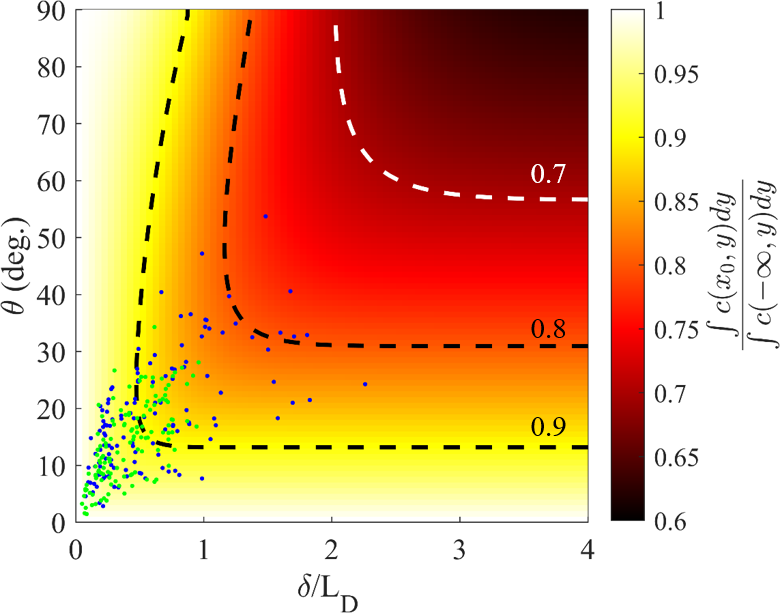}
\caption{The effect of topography on relative impurity dose at $t/t_D=1$ in the finite impurity source model. The blue and green points are the measurements from Fig. \ref{fig3} and Fig. \ref{fig6} respectively.}
\label{ThetaDeltaLDSpaceFiniteOSource}
\end{figure}

To understand the spatial homogeneity of impurities, it is important to compare the geometrically modified impurity profile with that far from the defect. A representative spatial distribution of impurities is shown in Fig. \ref{DiffusionComparison} (a). Taking a line profile through Fig. \ref{DiffusionComparison} (a), shown in gray, the normalized concentration far from the defect and at the bottom of the defect is plotted in Fig. \ref{DiffusionComparison} (b). Fig. \ref{DiffusionComparison} (b) shows that a substantial decrease in impurity content is possible at the vortex penetration position if the slope angle and step height are comparable or greater than the impurity diffusion length. This highlights that minimizing surface roughness is essential for a uniform introduction of impurities. The effect of topography on the relative impurity dose between the bottom of the sloped step located at $x_0$ and the flat surface far from the defect, $\int c(x_0,y)\,dy/\int c(-\infty,y)\,dy$, is shown in Fig. \ref{ThetaDeltaLDSpaceFiniteOSource}. The measurements from Fig. \ref{fig3} and Fig. \ref{fig6} are plotted within Fig. \ref{ThetaDeltaLDSpaceFiniteOSource} using a normalized step height ($\delta/L_D$) where $L_D$ for oxygen diffusion was estimated to be 30 nm for low temperature baked Nb \cite{romanenko2019SIMSOfHighQGradTreatments}. One can immediately see that low slope angles and step heights serve to topographically protect the uniformity of impurities, especially when the diffusion length is comparable to or less than the intergranular step height. This may be why rough surfaces, like that of buffered chemical polished Nb surfaces, are less responsive to processes like low-temperature baking or nitrogen infusion \cite{grassellino2017unprecedented} because their effective oxygen content is reduced since the topographic defect size is very large compared to the diffusion length. These impurity alloying processes are known to introduce impurities on a length scale comparable to the RF active region and the intergranular step heights described here \cite{Koufalis2017InfusionWhatsReallyGoingOn,bafia2021RoleOfOInHighGrad}. The effect of topography on relative impurity dose is qualitatively similar between all diffusion scenarios, as shown in the Supplementary Material.

\section{Conclusion}

Using WLI and AFM, we investigated the grain boundary defects resulting from electropolishing on bulk Nb samples. In particular, we characterized the slope angles, and intergrain steps introduced during electropolishing and calculated the associated magnetic field enhancement and superheating field suppression factors within the London theory. The results show that the grain boundaries host substantial slope angles and steps that compromise the high-field stability of the Meissner state. The estimated magnetic field enhancement and superheating field suppression factors present significant limitations on the electropolished bulk Nb for achieving peak fields near the superheating field. The measurements and analysis presented here show that the topographic defects hosted by electropolished Nb can account for a much larger degradation of the achievable peak magnetic field, $B_{max}=(\eta/\beta)B_v$, than previously expected based on prior measurements \cite{kubo2015field}. In clean Nb these defects can contribute field degradation factors $\eta \approx 0.8$ and $1/\beta \approx 0.8$ or a combined degradation, possible at grain boundary triple junctions, of $\eta/\beta\approx 0.64$. Considering an estimate of the superheating field of $\approx 210~\text{mT}$ (50 MV/m in TESLA-shaped \cite{AuneSuperconducting2000} cavities) these defects can account for a degradation to $B_{max}\approx 170$ mT ($40\, \text{MV/m}$) or a combined degradation $B_{max} \approx 140$ mT (32 MV/m) in some locations. This highlights the need for smoother surfaces to achieve field limits near or beyond the superheating field of Nb. Defects like these, if present in next-generation SRF cavity materials \cite{valente2016superconducting} or thin film superconductor-insulator-superconductor surfaces, should substantially alter their peak performance and optimal coating parameters \cite{KuboMultilayerReview2016}.

Electropolishing, despite hosting large slope angles, ultimately produces intergranular step heights comparable in length to the London penetration depth and diffusion length of heat treatments like low-temperature baking and nitrogen infusion. Intergranular step heights of this size or smaller are essential to preserve the relative impurity dose near topographic defects with large interior angles where vortex penetration is preferred. Ensuring the uniformity of the impurity concentration should enhance the effectiveness of shallow impurity profile surface preparation techniques like low-temperature baking or N-infusion in either the vortex nucleation or hydride precipitation scenarios. This reduction in intergranular step height provides a natural explanation for the difference in response of BCP'd and EP'd cavities to low-temperature baking, since the O source is better preserved for the electropolished surface. Interestingly, recent works \cite{Parajuli2024MagnetoThermalLimits,Howard2023HFQS} report a form of high-field Q-slope that remains unchanged by holding cavities between 90-150 K, a temperature range expected to encourage hydride growth and exacerbate high-field Q-slope \cite{Barkov2013HydridesAfterTreatments}. These observations may favor the vortex penetration scenario. 

Looking ahead to future particle accelerator requirements, reliably attaining accelerating gradients at 35 MV/m or higher in TESLA-shaped cavities is a major goal \cite{belomestnykh2022key} as well as the need for developing less hazardous surface processing techniques. There are several promising candidate polishing processes \cite{jmmp7020062, hryhorenko:srf19-thp002, chyhyrynets:srf23-mopmb009,hryhorenko:srf23-weixa06, pira:srf23-wecaa01}, but those processes must demonstrate superior surface finish if high-gradient performance is to be achieved. A term frequently used to describe the surface finish of electropolished Nb is "mirror-smooth". This by-eye evaluation of the surface finish is inadequate for high-field performance near the superheating field of Nb cavities and even less adequate for next-generation materials. We hope these measurements and analyses motivate the adoption of microscopic theories of superconductivity to more accurately predict how these defects and the impurity diffusion around them impact high-field RF performance.

\section{Acknowledgments}
The authors acknowledge the William \& Mary Applied Research Center for instrumentation resources and technical support. This was coauthored by Jefferson Science Associates LLC under U.S. DOE Contract No. DE-AC05-06OR23177. This material is based on the work supported by the U.S. Department of Science, Office of Science, Office of Nuclear Physics Early Career Award to A.M. Valente-Feliciano. The authors gratefully acknowledge C.E. Reece and C.Z. Antoine for useful discussions.

\bibliographystyle{apsrev4-1.bst}
\bibliography{main}
\beginsupplement

\end{document}